# Mapping the wavefunction of transition metal acceptor states in the GaAs surface

Anthony Richardella[1,2], Dale Kitchen[1,2,*], and Ali Yazdani[1]

1. Department of Physics, Joseph Henry Laboratories, Princeton University, Princeton, New Jersey 08544, USA.

2. Department of Physics, University of Illinois at Urbana-Champaign, Urbana, Illinois 61801, USA.

* Currently at: Milliken Research Corporation, PO Box 1927, M-405, Spartanburg, SC 29304

We utilize a single atom substitution technique with spectroscopic imaging in a scanning tunneling microscope (STM) to visualize the anisotropic spatial structure of magnetic and non-magnetic transition metal acceptor states in the GaAs (110) surface. The character of the defect states play a critical role in the properties of the semiconductor, the localization of the states influencing such things as the onset of the metal-insulator transition, and in dilute magnetic semiconductors the mechanism and strength of magnetic interactions that lead to the emergence of ferromagnetism. We study these states in the GaAs surface finding remarkable similarities between the shape of the acceptor state wavefunction for Mn, Fe, Co and Zn dopants, which is determined by the GaAs host and is generally reproduced by tight binding calculations of Mn in bulk GaAs [Tang, J.M. & Flatte, M.E., Phys. Rev. Lett. 92, 047201 (2004)]. The similarities originate from the antibonding nature of the acceptor states that arise from the hybridization of the impurity d-levels with the host. A second deeper in-gap state is also observed for Fe and Co that can be explained by the symmetry breaking of the surface.

# I. Introduction

A dilute magnetic semiconductor (DMS) is a semiconductor doped with a random distribution of magnetic, usually transition metal, atoms whose local moments can couple, giving rise to a collective ferromagnetic state.[1] In the prototypical DMS, $Ga_{1-x}Mn_xAs$, Mn dopants are acceptors providing holes that are believed to be responsible for mediating the ferromagnetic interaction between Mn $3d^5$ core spins. The nature of the interaction between the hole and the core spin is then a central question whose answer hinges on the character of the hole state. Transition metals substitute the trivalent cation in III-V materials. In GaAs, Zn forms a shallow acceptor whose valence band-like hole state is well described by effective mass theory. Fe and Co are deep acceptors that are expected to have a localized state that is largely determined by their atomic central cell potential. Mn sits between these cases as a fairly deep acceptor that binds a hole in an $Mn^{2+}3d^5$+hole configuration.[2] In this sense it is like Zn, but its binding energy is larger because there is hybridization between the d levels and the host which, importantly, provides a means of coupling to the spin of the core state. The strength and influence of this hybridization on the nature of the hole state has been a source of debate. In the effective mass limit, when the doping crosses the metal-insulator transition, the holes reside in the valence band and their interaction mediating ferromagnetism between Mn sites can be understood within the RKKY model.[3, 4] Others have questioned this, finding the hybridization to be large enough to give the hole significant impurity-like d character and the magnetic order to arise from exchange interactions between localized Mn impurity states[5, 6]. The two limits will in general predict different anisotropies in both the defect state wavefunctions and the magnetic interaction and it will be necessary to understand these to both to raise $T_c$ and to effectively make devices that utilize these materials.[7] Thus the question of the nature and character of the acceptor states in III-V materials is an important one and one that can be probed by comparing the acceptor states of various substitutional transition metal impurities. To do this we have used a scanning tunneling microscope (STM) and an atomic scale manipulation technique to substitute transition metal atoms into a GaAs host.

## *II. Experiment*

The DMSs are complex materials to study experimentally due to a number of factors such as the high doping concentrations, compensating defects due to non-equilibrium growth conditions, and disorder. This challenges the ability of macroscopic scale measurements that average over large areas to provide insights into the microscopic scale interactions that give rise to ferromagnetic order. Alternatively, microscopic visualization of Mn in the surface of GaAs has been used to probe ferromagnetic interactions in these materials.[8] Using an STM, various species of isolated acceptor defects can be created and probed in an identical layer, allowing spectroscopic resonances, unambiguously associated with the defect wavefunctions, to be directly compared.

The experiments were performed using a home-built cryogenic STM that operates near 4 K in ultra-high-vacuum. Wafers of p-type GaAs, doped at $10^{19}$ Zn atoms $cm^{-3}$, were cleaved *in situ* to expose a (110) surface. The Zn dopant concentration was measured in STM topographs of the GaAs (110) surfaces prior to depositing foreign atoms. In addition to Zn below the surface, a defect in the surface matching the expected concentration of Zn dopants was observed, which we identify as Zn in the surface layer. This added the additional effective mass-like, non-magnetic transition metal acceptor in the same configuration to the other magnetic atoms studied. We evaporated small concentrations of Mn, Fe, or Co atoms (~0.5% monolayer) from *in situ* sources onto the cold GaAs surface. The various species could be distinguished between samples in filled state topography. Previous STM studies of metal atoms on GaAs (110) deposited enough atoms to form clusters.[9-12] In contrast, the adsorbate atoms were clearly identifiable as isolated atoms on the surface and no Fermi level pinning was observed.

The STM atomic manipulation technique involved placing the STM tip above the adsorbate atom and sweeping the tunneling bias from negative to positive voltage across the GaAs energy gap with the feedback loop off, tunneling energetic electrons into the adatom. Doing so we find that Mn, Fe and Ga adatoms can be made to move randomly on the surface within a half circle centered about the [001] direction, while Co atoms move oppositely. Using higher current and voltages, the transition metal atoms could be made to substitute into Ga sites in the first layer of the (110) surface, ejecting the Ga atom to the surface. The energetics of the substitution process varied between species but were generally similar. Using a setup condition of 200pA at -1.5V, the tunnel current would drop to zero as the bias voltage was swept through the gap, begin to rise sharply near +1V and then drop suddenly a few hundred mV later when the substitution occurred, providing a rough qualitative measure of the activation barrier required to be overcome to occupy the cation site. Only Co could be induced to swap with a Ga adatom and return to the Co adatom configuration again, as shown in Fig. 1. After substitution, the displaced Ga atom often remained loosely bound to the newly created defect site, proving more difficult to remove for Co than for Fe or Mn. The substitution process is likely similar to the ejection of cation atoms to the surface that can occur during the initial growth of transition metal overlayers on III-V surfaces at room temperature, suggesting this technique may work on a variety of substrates.[13, 14] Recent work on the deposition of Fe films on GaAs that found that the intermixing of species and formation of compounds at the interface can be significantly reduced by low temperature deposition[15, 16] is potentially consistent with our observation that the adatoms at low temperature sit on the surface as adatoms and do not substitute with atoms of the substrate until we manipulate them.

Topography of Mn, Zn, Co, and Fe in the surface layer are displayed in Fig. 2. The topography maps the acceptor states by applying the bias voltage near the conduction band edge so that the image is the sum of all states in the gap above the Fermi energy. An anisotropic star-shaped structure is apparent for each with $C_s$ symmetry similar to that observed for subsurface acceptors in GaAs.[17-19] The wavefunction of Mn in bulk GaAs predicted by tight-binding calculations viewed in the plane of the Mn site resembles our data well.[8, 20] It is known that the depth of the dopant beneath the surface affects the symmetry of the acceptor wavefunction observed by STM, with Mn going from $C_{2v}$ to $C_s$ symmetry the nearer to the surface it is, possibly as a result of surface related strain.[21] The wavefunctions in Fig. 2 can be viewed as the extreme limit of that trend. Recent tight-binding work that attempted to take the surface into account is consistent with this trend but doesn't include the case of Mn at the surface.[22] The wavefunctions can be described as a central lobe with arms along the $[1\bar{1}0]$ directions and legs along $[1\bar{1}1]$. The dominant half of the subsurface Mn bowtie wavefunction is associated with the central lobe, which expands in the $[00\bar{1}]$ direction when located in layers beneath the surface, while the weaker half arises from the leg features. The arms are not observed in subsurface Mn in GaAs but a similar feature has been observed for Cd in GaP.[23] The arm and leg features of the Zn impurity are weaker than Mn, suggestive of the triangular wavefunction observed for subsurface Zn dopants.[18] Whereas the triangular shape of subsurface Zn dopants observed by STM has been associated with multiple resonances that are observed throughout the band gap,[24-26] the spatial structure of a Zn in the surface layer is clearly identified with a distinct single state in the dI/dV density of states spectra. The observed wavefunction is measured as a wide in-gap state in spectroscopy peaked near 0.65 eV for the surface Zn acceptor and at 0.85eV for Mn, shown in Fig. 3(a).

While the shape of the Mn and Zn acceptors in the topographs directly correlate with single, strong resonant levels, Fe and Co acceptors have more complicated spectral signatures. The Fe and Co states show increased spectral weight on the arm and leg features. Spectroscopy shows that both Fe and Co have a low energy state close to the Mn level at 0.87 and 0.92 eV, and a higher energy state at 1.5 and 1.15 eV respectively , shown in Fig. 3(b) and Fig. 3(c). Spatial maps of the energy-resolved density of states in Fig. 4 show that the shape of the lower energy state is similar to that of Mn. The higher energy state is responsible for the greatly enhanced arm-like features. The low contrast at the location of the dopant in the maps of the higher energy states in Fig. 4 is somewhat exaggerated due to the lower energy state causing the STM tip to move out, decreasing the sensitivity. Nevertheless, the maps clearly show that spatial dependence is quite different for the two states.

The filled state topography of the defects (insets Fig. 2) shows an enhancement on the two in-row $[1\bar{1}1]$ nearest-neighbor As sites, with some minor variation between the species. The lack of any depression in the valence band topography of the p-type GaAs around the impurities shows they are not positively charged and thus the states in the gap are acceptor like. At biases close the Fermi energy we observe an apparent lattice distortion involving the nearest neighbor As atoms for all species except Co. This is most clearly seen with Mn, where at low bias one As appears to buckle upward and another down. Electrons tunneling inelastically with sufficient energy can cause the distortion to switch to the other equivalent configuration, with the down As now buckling upward and vice versa as seen in (b) and (c) of Fig. 5. Larger biases cause the switching to occur more rapidly resulting in noise in the topography around the feature, similar to that observed due to switching behavior in a variety of other systems with STM.[27, 28] At large biases the switching occurs quickly enough that only the average is observed[29], as in the insets of Fig. 2. Coupling to phonons can provide an efficient means of dissipating energy during the capture and recombination process of carriers trapped by deep levels.[30] A similar multiphonon emission process could explain the large magnitude of the dI/dV signal when tunneling into these impurities deep in the gap as well as for their width in energy.

## *III. Discussion*

While the similarity of the ground state of all four species of acceptor may at first be surprising, it follows from the well known effect of symmetry on the hybridization of a substitutional impurity with the host. The defect states can be seen as arising from the interaction of the atomic states of the isolated impurity with the dangling bond states of an ideal vacancy in the lattice (Fig. 6). Clearly, if the vacancy is filled with the host atom the hybridization forms a bonding state in the valence band and an antibonding one in the conduction band. It is the difference in the hybridization that occurs with an impurity that gives rise to the defect states.[31] The case of Zn is the easiest to understand: the Zn d levels are deeply bound below the valence band edge and non-interacting and the Zn valence orbitals are comparable to but higher than those of Ga. This gives rise to a state with primarily host-like bonding character that isn't pushed all the way down into the valence band and an antibonding state with impurity character above the conduction band edge.

As shown in Fig. 6, the case when the d levels are interacting is more involved and is well summarized by Mahadevan and Zunger.[5] The s and p levels that make up the $sp^3$ hybrids of the vacancy transform under zinc-blende $T_d$ symmetry according to $A_1$ and $T_2$ irreducible representations. The $a_1$ state is a singly degenerate s-like state, while the $t_2$ is triply degenerate p-like. Before hybridization, the $t_2$ levels of the Ga vacancy are near the valence band edge. Under the same symmetry, the d-levels of the transition metal are crystal field split into an e symmetric doublet and $t_2$ symmetric triplet. The $t_2$ states of the impurity and vacancy will strongly hybridize into bonding and antibonding states while the e states do not. Mahadevan and Zunger calculate that the impurity d levels of Mn, Fe and Co lie energetically beneath the vacancy level. Therefore, the $t_2$ derived in-gap states are antibonding and predominantly host-like, though the stronger the hybridization the more mixed character they will have. This is the origin of the $Mn^{2+}3d^5$ + hole configuration, the hole is host-like and, though antibonding instead of bonding, it shares the same $t_2$ (p-like) symmetry as the valence band. The core spin in preserved in the deep impurity-like e and bonding $t_2$ states that fill in a high spin configuration according to Hund's rules.

With this in mind, a few conclusions can be draw. First, the anisotropy of the ground state for all four acceptors is similar, implying they are all states of the same ($t_2$) symmetry. It should be noted that the character of acceptor state will depend sensitively on the order of levels in energy and be determined by the lowest unoccupied state. The similarity between them is most simply explained if the bonding d levels fill according to their atomic configuration. Second, the spatial extent of the Mn, Fe and Co ground state is the same and that of Zn is smaller. This is a result of the antibonding nature of the Mn, Fe and Co states causing the wavefunction to be more delocalized onto the As dangling bonds than the bonding nature of the Zn state. This result is somewhat counterintuitive, because as the shallowest acceptor the Zn wavefunction is expected to be the most delocalized. It arises from the fact that STM doesn't sample the wavefunction within the material but rather the part that is exponentially decaying into the vacuum at the location of the tip. Since the acceptors at the surface have no layers above them for the wavefunction to spread out in, instead the STM only images the core of the wavefunction at the impurity site. Third, a partially filled triply degenerate $t_2$ level could be expected to be susceptible to Jahn-Teller distortions as are observed for some configuration terms of Fe and Co in III-V materials, potentially explaining Fig. 5.[32, 33]

In our analysis thus far, however, we have ignored the effect of the surface and surface reconstruction which lowers the symmetry from $T_d$ all the way to $C_s$ ((1$\bar{1}$0) mirror plane). This will split the $t_2$ states into functions of parity, creating two a' (even) levels and one a'' (odd) level (Fig. 6). While it is hard to estimate the resulting energies of these levels, ab-initio calculations of the unrelaxed surface Ga vacancy in GaP found the $t_2$ level to split into one a' level in the valence band and the remaining a' and a'' levels to be nearly degenerate above the valence band maximum.[34] This implies the possibility of two states in the gap, one having even symmetry and other odd. The comparison to the spatial structure of the two states of Fe and Co is immediately obvious: the lower energy state being symmetric with respect to the (1$\bar{1}$0) plane and the higher energy state being consistent with the wavefunction squared of an odd state with respect to the mirror plane. For the higher state, it is unlikely that the doubly ionized state is being observed, but rather that the electron is tunneling into an unoccupied excited state. In Fig. 6(c) one possible level order that allows for two unoccupied states of different symmetry is shown.

Finally, we attempt to address the energies of the states. The order of the acceptors in bulk is Zn (31 meV), Mn (113 meV), Co (160 meV) and Fe(~0.5 eV).[35] At the surface we observe the same trend with the exception that the ground state of Fe is below Co. The shift to higher energy is likely due both to tip induced band-bending and the energy of the acceptor states being affected by being at the surface. If the Ga vacancy dangling bond levels are higher at the surface than in bulk this would lead directly to a deeper acceptor state. It is worth noting also that charge transfer from cation to anion sites due to the surface reconstruction is known to increase the binding energy of the surface Ga 3d levels by ~0.3 eV.[36, 37] If this applies to the transition metal impurities, it would be expected to push the d-states deeper into the valence band lessening the p-d hybridization, and pinning the acceptor energies closer to the Ga vacancy dangling bond energy, resulting in the confluence of similar energies for the first states of Mn, Fe and Co. The energy splitting of the states may also be correlated with the amount of lattice distortion observed for each impurity, with Co having no observable distortion and the smallest energy splitting and Fe having more distortion and splitting. As Mn had the most distortion, the upper energy level would be pushed into the valence band. This would explain the lack of an observation of a second level for Mn which would be expected from the preceding analysis of the origin of the acceptor states.

## *IV. Conclusion*

We have demonstrated an atomic scale technique to study single transition metal impurities in the surface of a III-V semiconductor, successfully substituting single Mn, Fe, and Co atoms into the GaAs (110) surface, as well as identifying native Zn dopants in the same layer. The similarity of the anisotropic wavefunctions measured for all these acceptors shows the importance of the host in determining the shape of these states even for deep non-effective mass acceptors. The character of these states can be seen to follow directly from simple symmetry arguments and the hybridization of the impurity and host states.

This work was supported by ARO W911NF-07-1-0125, ONR N00014-07-1-0348 and NSF DMR-0704314.

## *Figure Legends*

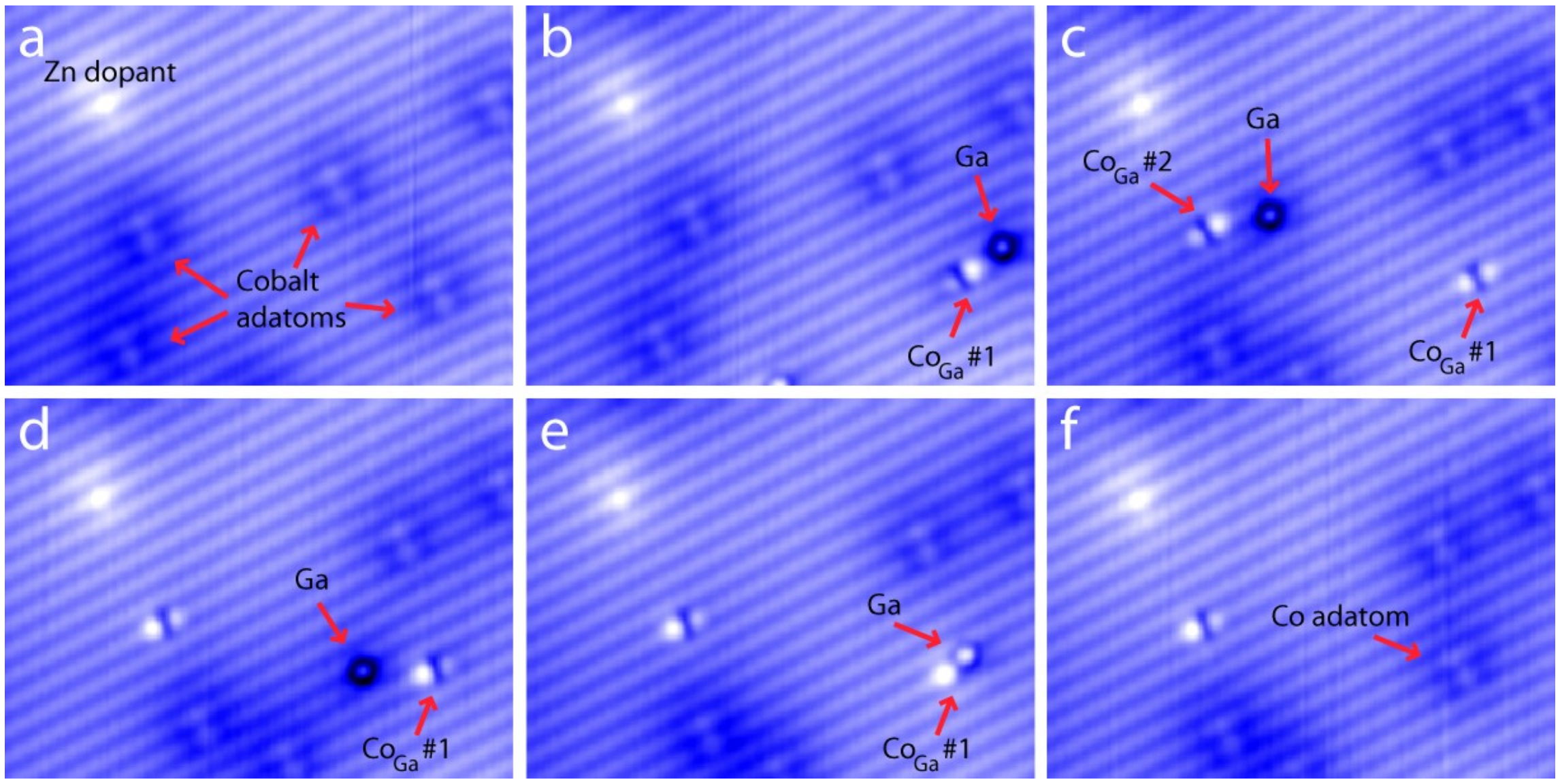

**FIG. 1 (color).** Two substitutions of cobalt adatoms into the lattice showing cobalt can de-substitute and come back to the surface. **(a)** Several cobalt adatoms that appear as depressions on the lattice. A subsurface Zn dopant can be seen in the upper left. **(b)** A voltage pulse from the STM tip, as described in the text, causes a cobalt atom in the lower right to substitute into the lattice, kicking a gallium atom to the surface. **(c)** The gallium from the first site is moved out of view and a second cobalt is substituted at the center. **(d,e)** The gallium adatom from the second site is manipulated to the location of the first site. **(f)** A voltage pulse causes the gallium to go in and the cobalt comes back to the surface. (All images are 120Å x 90Å at -1.5V)

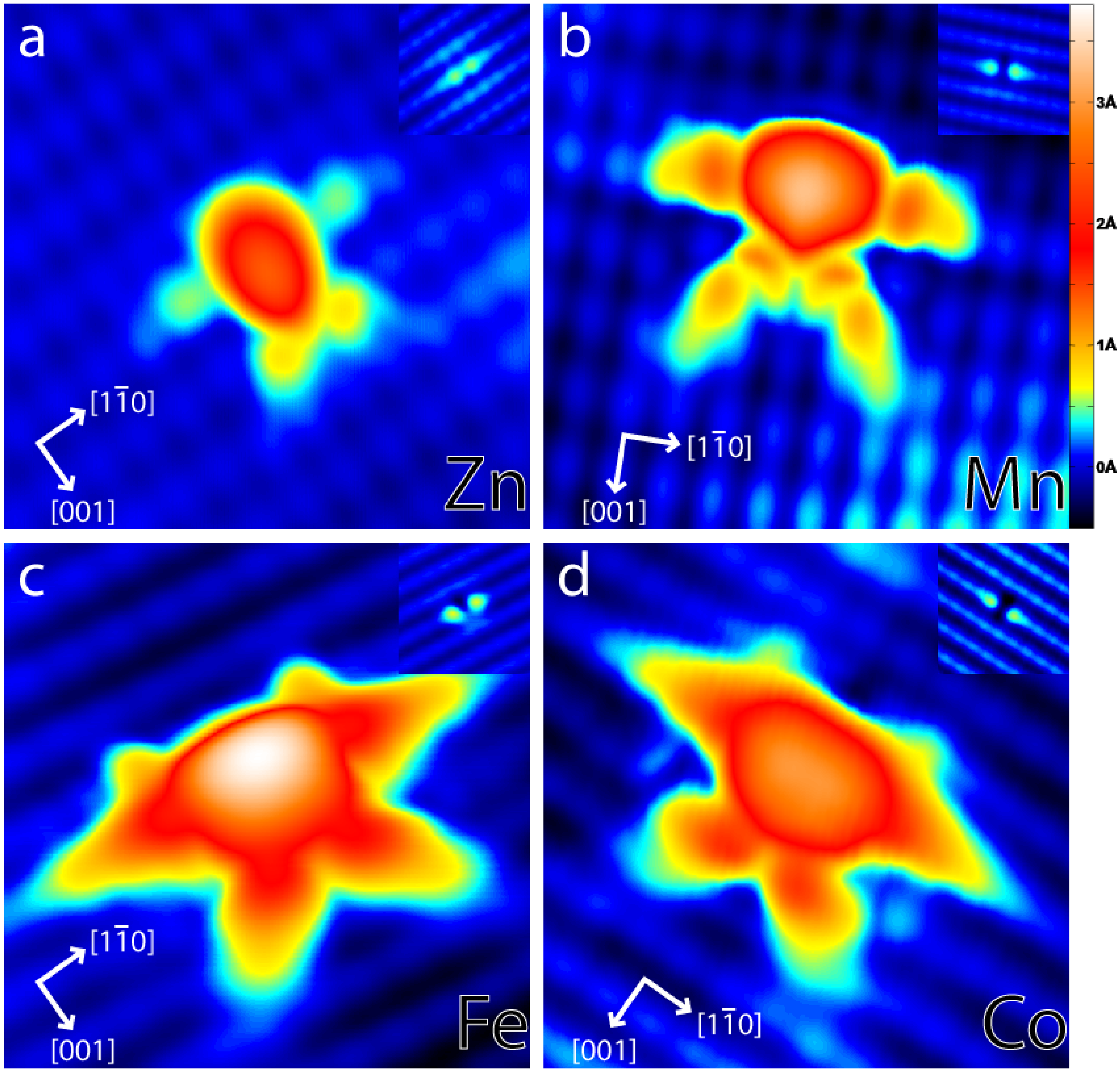

**FIG. 2 (color).** Topography of in-gap states (40Å by 40Å) and insets of valence band states (30Å by 30Å). Topography in insets has been multiplied by two to enhance the contrast. **(a)** Zn (1.6V, inset: -1.4V), **(b)** Mn (1.6V, inset: -1.3V) , **(c)** Fe (1.5V, inset: -1V) **(d)** Co (1.5V, inset: -1.3V)

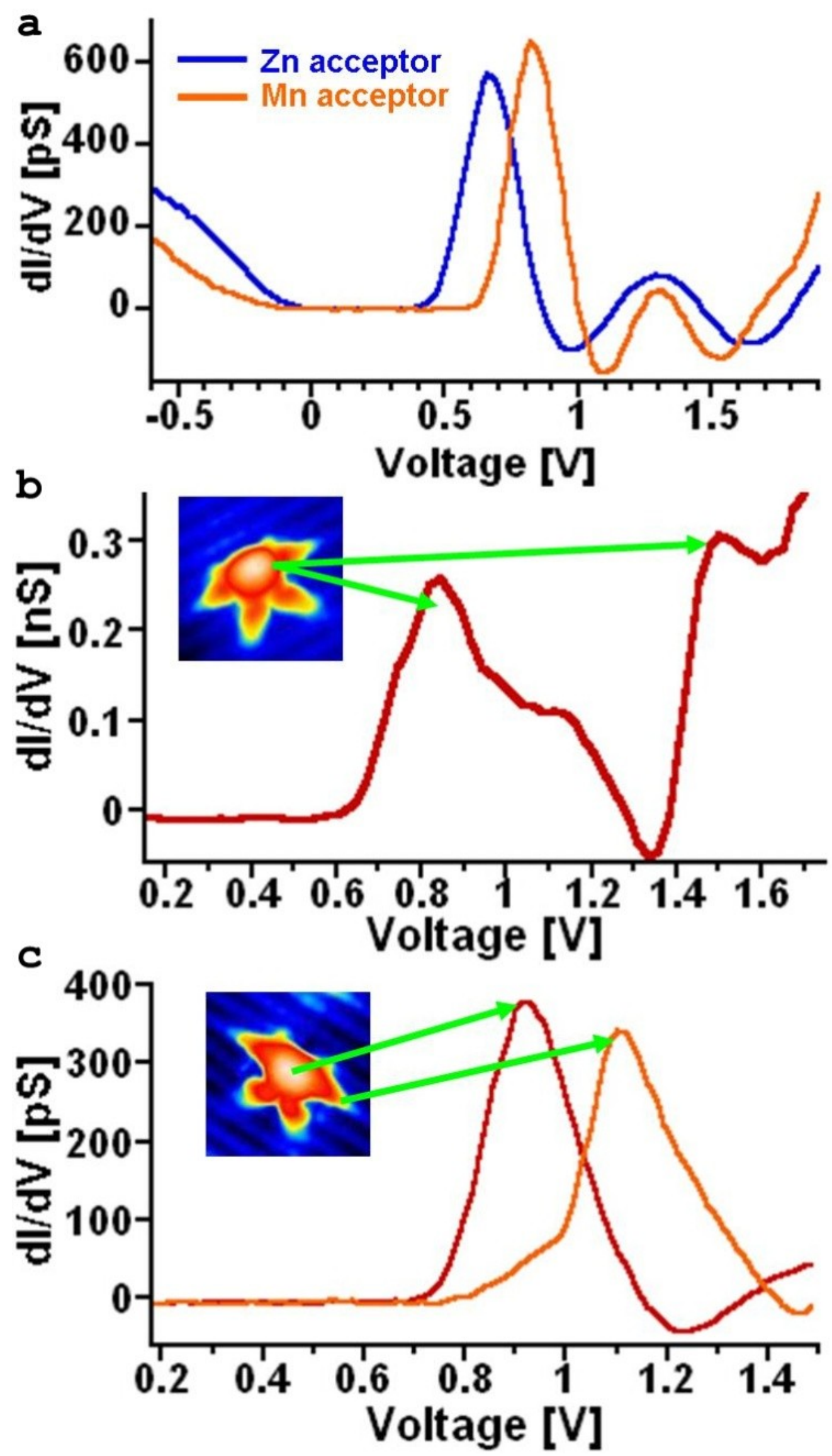

**FIG. 3 (color). (a)** dI/dV measurements over an Mn acceptor and Zn acceptor; each show a strong, broad in-gap resonance. The Zn resonance peaks near 0.65±0.05 V and the Mn resonance peaks near 0.85±0.03 V. **(b)** dI/dV measurements near an Fe acceptor, with the tip between head and arm features (see inset), bring out two states near 0.87±0.05 V and 1.52±0.05 V. **(c)** Two dI/dV measurements taken near a Co acceptor show two distinct states, with the tip centered over the $Co_{Ga}$ a lower energy resonance near 0.92±0.05 V, and the tip over the arm feature a higher energy peak near 1.15±0.05 V.

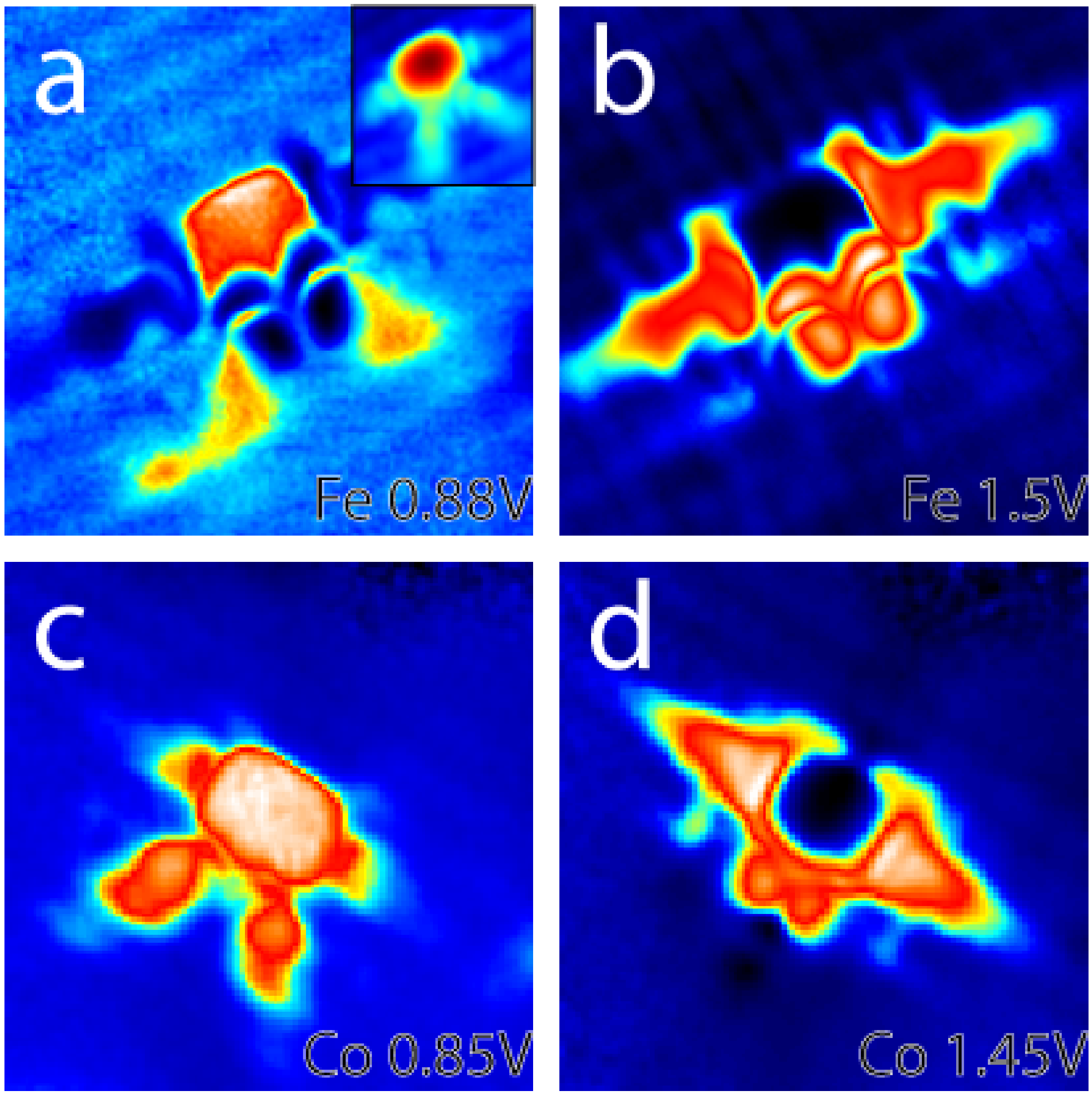

**FIG. 4 (color).** Differential conductance maps of Fe and Co acceptors (50Å by 50Å). Energy maps near the $Fe_{Ga}$ show **(a)** the lower resonance state mapped at 0.88 V (inset: 30Å by 30Å topography taken at 1V better shows spatial extent of the state), and **(b)** the higher energy resonance mapped at 1.50 V. Energy maps of an isolated $Co_{Ga}$ **(c)** near the lower resonance peak energy of 0.85 V and **(d)** near the higher energy resonance of 1.15 V with a setup voltage of 1.05 V. The $Fe_{Ga}$ setup voltage was 1.45 V which created a larger shadow effect in (a) than in (c) since the set voltage for the $Fe_{Ga}$ map was nearer its high energy resonance. Both (b) and (d) exaggerate the depression in the center because the tip is further from the surface due to the tunneling contribution from the lower energy resonance.

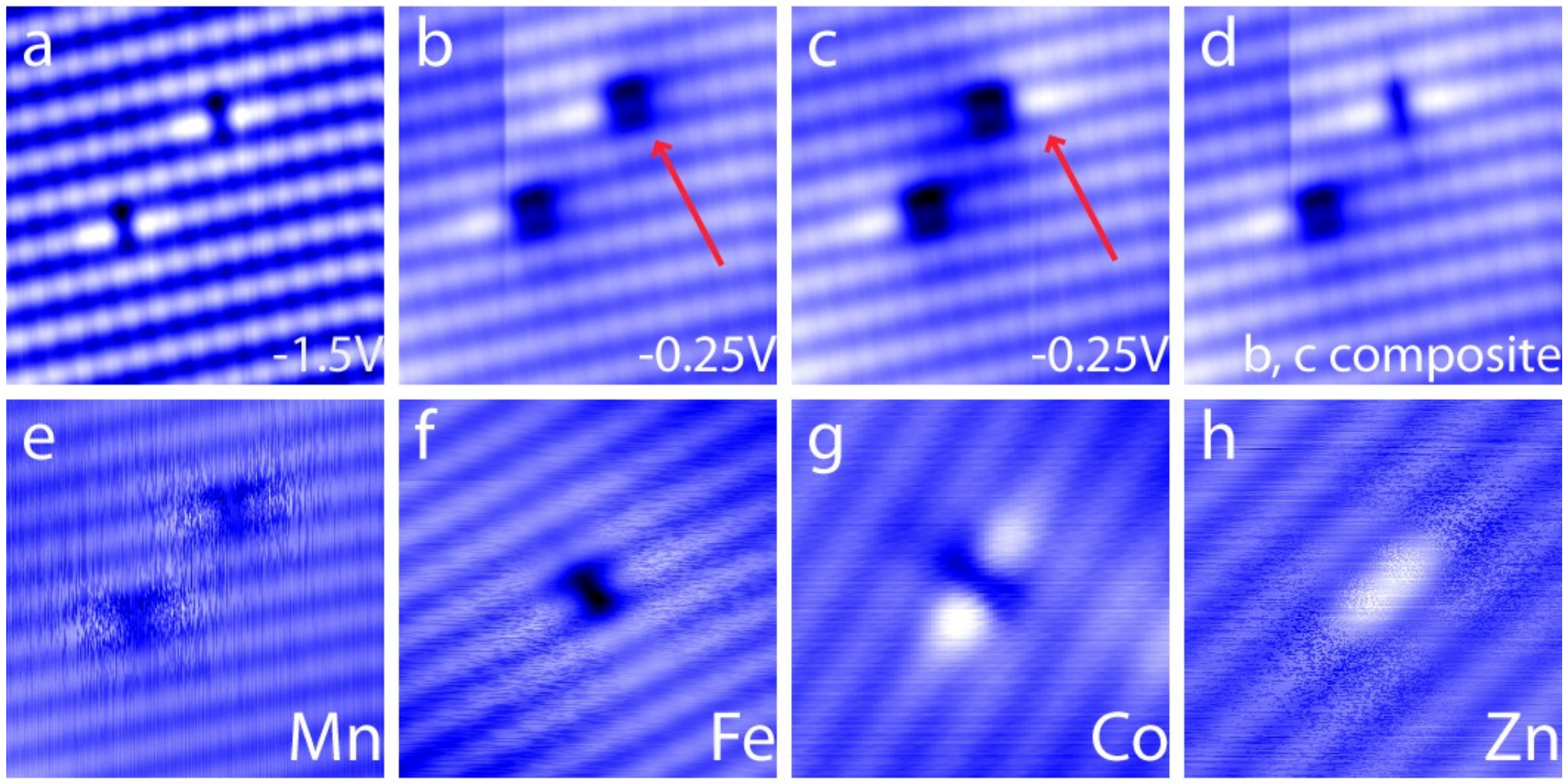

**FIG. 5(color). (a)** Valence band topography of two $Mn_{Ga}$ sites showing an enhancement of the adjacent arsenic atoms. (50Å by 50Å, -1.5V) **(b)** Low bias topography of the same area shows an enhancement on only one arsenic and depression. Scan proceeds from right to left. A glitch occurs ¾ of the way through the measurement (-0.25V). **(c)** Result of glitch in (b), feature has switched to the mirror configuration. **(d)** Composite image made by combining (c) and (d) showing how the high bias image is a mixture of the two configurations. **(e)** As the bias is increased the switching rate increases, appearing as noise in the topography. (-0.5V) **(f-h)** Noise patterns were observed around (f) $Fe_{Ga}$ (45Å by 45Å, -0.3V) and (h) $Zn_{Ga}$ (32Å by 32Å, -0.5V) but not (g) $Co_{Ga}$ (32Å by 32Å, -0.3V).

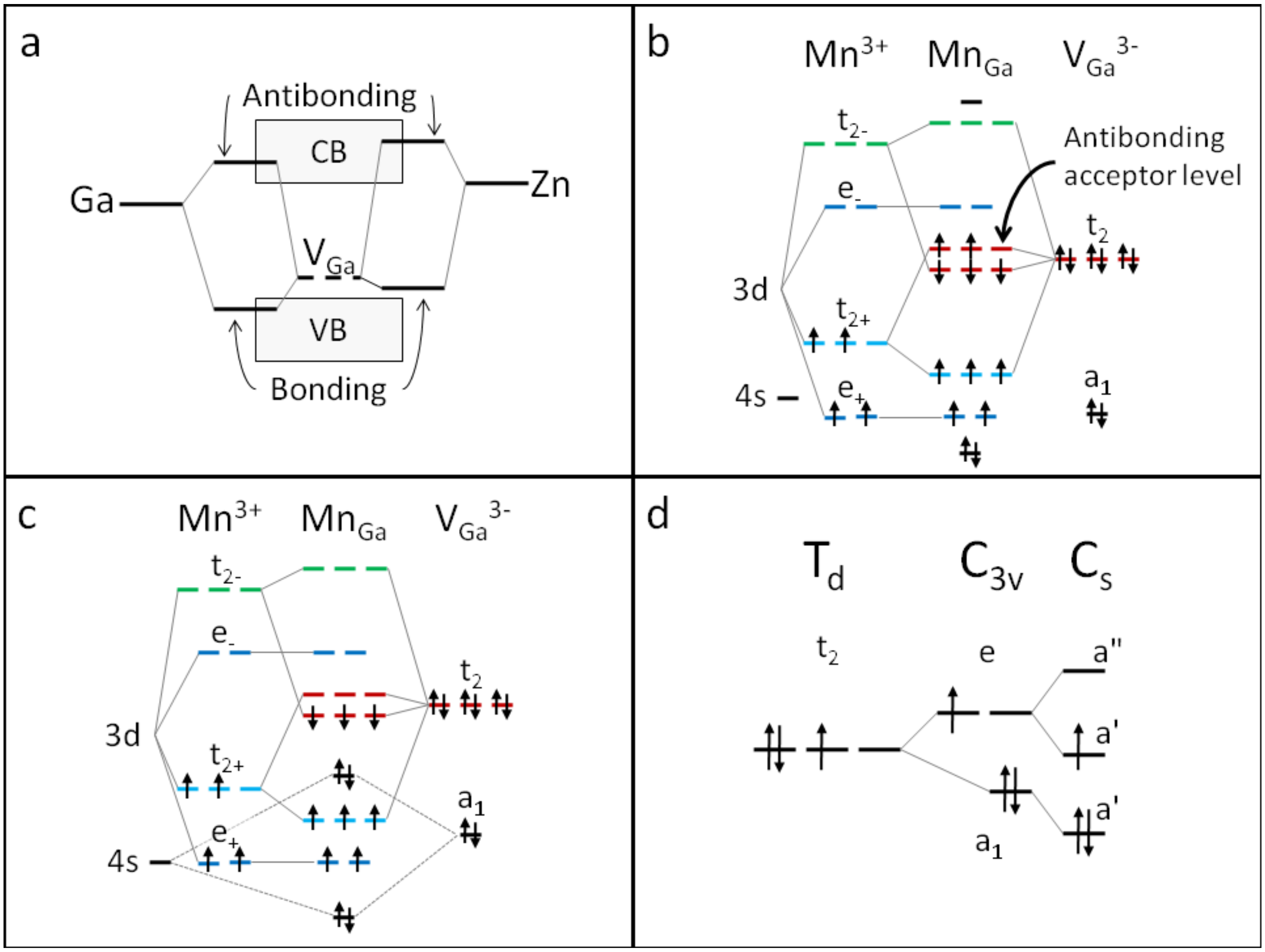

**FIG. 6(color). (a)** Schematic of hybridization of an atom with an ideal gallium vacancy ($V_{Ga}$) leading to a shallow acceptor level such as for Zn. The valence levels of the impurity are higher in energy than the host atom resulting in a state just above the band edge that shares the valence band's bonding character. **(b)**Hybridization of levels giving rise to the Mn acceptor following the model of Ref 5. In contrast to (a) the acceptor is antibonding in character. States are labeled +(-) for majority(minority) spin. Colors represent the initial state the hybridized $Mn_{Ga}$ level most resembles in character. **(c)** Possible level ordering that would give rise to a symmetric acceptor state for an Mn under the influence of the symmetry breaking of the surface, as described below. **(d)** Splitting of a $t_2$ state by the surface. Removal of one neighboring As atom lowers the $T_d$ symmetry to $C_{3v}$, the rest of the missing atoms at the surface and the GaAs surface reconstruction lower the symmetry further to $C_s$ ((1$\bar{1}$0) mirror plane). For there to be a symmetric (a') acceptor level at the surface there can be at most three electrons in the vacancy-like $t_2$ level. For Fe and Co, this would require that their additional electrons fill the impurity-like e and $t_{2}$- levels according to their atomic configuration.